\documentclass[11pt, draftclsnofoot, onecolumn]{IEEEtran}

\IEEEoverridecommandlockouts
\usepackage{cite}
\usepackage{xcolor}
\usepackage{amsmath,amssymb,amsfonts}
\usepackage{algorithmic}
\usepackage{comment}
\usepackage{graphicx}
\usepackage[ruled,vlined]{algorithm2e}
\usepackage{booktabs}
\usepackage{subfig}
\usepackage{soul}
\usepackage{mathtools} 
\usepackage{wrapfig}
\allowdisplaybreaks
\usepackage{bigdelim}
\usepackage{caption,setspace}
\usepackage{textcomp}

\DeclareMathOperator*{\argmin}{arg\,min}
\newtheorem{prop}{Proposition}

\newtheorem{lemma}{Lemma}
\newtheorem{corollary}{Corollary}

\usepackage{xcolor}
\def\BibTeX{{\rm B\kern-.05em{\sc i\kern-.025em b}\kern-.08em
    T\kern-.1667em\lower.7ex\hbox{E}\kern-.125emX}}
\begin{document}

\title{Node Deployment in Heterogeneous Rayleigh Fading Sensor Networks  \\
\thanks{This work was supported
in part by the NSF Award CNS-2229467.}
}

\author{\IEEEauthorblockN{\vspace{10mm}Saeed Karimi-Bidhendi and Hamid Jafarkhani }
\IEEEauthorblockA{{\small \\ \vspace{-1mm} Center for Pervasive Communications and Computing, University of California, Irvine, CA, 92697} \\ \vspace{-1mm}
{\small Email: \{skarimib, hamidj\}@uci.edu}}
}

\maketitle

\begin{abstract}

We study a hierarchical heterogeneous Rayleigh fading wireless sensor network (WSN) in which sensor nodes surveil a region of interest (RoI) and use access points (APs) as relays to transmit their sensed information to base stations (BSs). By considering both large-scale path-loss signal attenuation and small-scale signal variation due to Rayleigh fading, we formulate the node deployment problem as an optimization problem intended to minimize the network's wireless communication power consumption. Given ergodic capacity constraints on all wireless links, we study the necessary conditions for an optimal AP and BS deployment. These necessary conditions are then assembled in the form of an iterative algorithm to deploy nodes. Finally, we establish the efficacy and superiority of our proposed node deployment algorithm against similar methods in the literature. 

\end{abstract}


\section{Introduction}\label{introduction}

Wireless sensor networks (WSNs) are primarily used to monitor physical phenomena such as temperature, humidity, barometric pressure,  etc. inside a region of interest (RoI) and transmit the sensed information back to base stations (BSs) \cite{yick2008wireless, zheng2009wireless}. Since battery replacement in harsh or hostile regions is impractical \cite{farsi2019deployment}, energy-efficiency is considered the most crucial factor for WSNs' continued operation \cite{matin2012overview}.  Energy consumption is largely dominated by the communication energy \cite{yousefi2004power,warrier2016energy}; thus, access points (APs) are often deployed to act as relays and facilitate data transfer from sensors to BSs \cite{eu2012adaptive}.

To improve WSNs' energy-efficiency, various methods have been proposed;  ranging from optimizing data routing \cite{benaddy2017mutlipath, nakas2020energy, mostafaei2018energy, sajwan2018hybrid} to effectively scheduling active and sleep modes for sensors \cite{wan2018energy, radhika2021fuzzy, karthihadevi2017sleep}. Among these methods, energy-efficient node deployment has received much attention  as the signal-to-noise ratio (SNR) at the receiver decreases as the distance between the transmitter and receiver increases. Based on the network's setup, some methods are executed in a centralized manner \cite{abo2016centralized} while others are distributed and only utilize local information at each node to achieve optimal deployment \cite{chatterjee2017load, aliyu2016coverage}. In addition, some methods are configured for static networks where nodes are placed at precalculated positions \cite{karimi2020energy,  guo2016sensor, karimi2021energy} while others consider mobile networks where each node moves from its initial position to its optimal location \cite{karimi2021energy, guo2019movement, abo2016centralized}. The optimal deployment is highly affected by nodes' characteristics. Homogeneous WSNs, in which network nodes have identical hardware properties such as antenna gain and sensitivity, are studied extensively \cite{benaddy2017mutlipath, cortes2005spatially, guo2018source}. Recently, we have studied the optimal deployment of heterogeneous WSNs, in which nodes have different characteristics and the optimal regions are often non-convex and even disconnected  \cite{karimi2020energy,  guo2016sensor, karimi2021energy}. 
While the heterogeneity of nodes is taken into account in \cite{karimi2020energy,  guo2016sensor, karimi2021energy}, these studies along with the majority of the work in the literature, ignore the real-world characteristics of the RoI and overlook the stochasticity of the communication channel due to the fading phenomena. Moreover, they employ a very simplistic radio energy model in which the exponential dependence of the required transmission energy on the rate is neglected. This in turn leads to underestimation of the actual energy consumption and substantially decreases the reliability and durability of these networks; thus, increasing the need for further research into the development of models that can realistically simulate real-world properties of WSNs.

The main motivation and key characteristics of this work is to address the aforementioned fundamental shortcomings. 
This manuscript studies the optimal deployment in heterogeneous Rayleigh fading sensor networks to minimize the wireless communication power consumption given ergodic capacity constraints on all wireless links.  We have aimed to consider the small-scale fading and the exponential dependence of the required transmission energy on the rate in heterogeneous WSNs, a shortcoming of the existing literature on this topic. Notably,  we consider a radio energy model that is based on both large-scale path-loss signal attenuation and small-scale signal variation due to Rayleigh fading, and takes the heterogeneity of nodes into account. By marginalizing the randomness of the channel capacity due to the Rayleigh fading, and considering the ergodic capacity for all wireless links, we derive the necessary conditions of optimal deployment and design an energy-efficient algorithm to deploy nodes.

The rest of the manuscript is organized as follows: The system model and problem formulation are outlined in Section \ref{system_model}. In Section \ref{Optimal_Deployment_under_Ergodic_Capacity_Assumption}, the optimal deployment in heterogeneous Rayleigh fading WSNs given ergodic capacity constraints on all wireless links is studied, and an iterative algorithm based on the obtained necessary conditions is provided. 
The proofs of the results in Section \ref{Optimal_Deployment_under_Ergodic_Capacity_Assumption} are provided in Appendices.
Finally, Section \ref{experiments} presents the simulation results and Section \ref{conclusion} concludes the paper.

\section{System Model and Problem Formulation}\label{system_model}

We study a heterogeneous WSN comprised of homogeneous sensors, $M$ heterogeneous BSs, and $N$ heterogeneous APs that act as relays and facilitate data transfer from sensors to BSs. The RoI $\Omega \subseteq \mathbb{R}^2$ is a convex polygon including its interior. The distribution of densely deployed sensors is characterized by a continuous and differentiable function $f:\Omega \longrightarrow \mathbb{R}^+$, i.e., $\int_W f(\omega)d\omega$ is the total number of sensors within $W\subseteq \Omega$. For applications such as surveillance sensor networks and traffic control systems that require continuous feed of visual and/or audio information, sensors transmit their data with the bit-rate $R_b$ where $R_b$ is a constant due to sensors' homogeneity \cite{guo2018source}. For applications in which sensors have sporadic activity, the assumption of constant bit-rate for homogeneous sensors can be adapted accordingly by considering time-division multiplexing for sensors' activity. Hence, $R_b\int_W f(\omega)d\omega$ denotes the amount of data generated by all sensors within the region $W$ in one time unit.

\begin{figure}[t]
\setlength\abovecaptionskip{0pt}
\setlength\belowcaptionskip{0pt}
\centering
\includegraphics[width=4.0in]{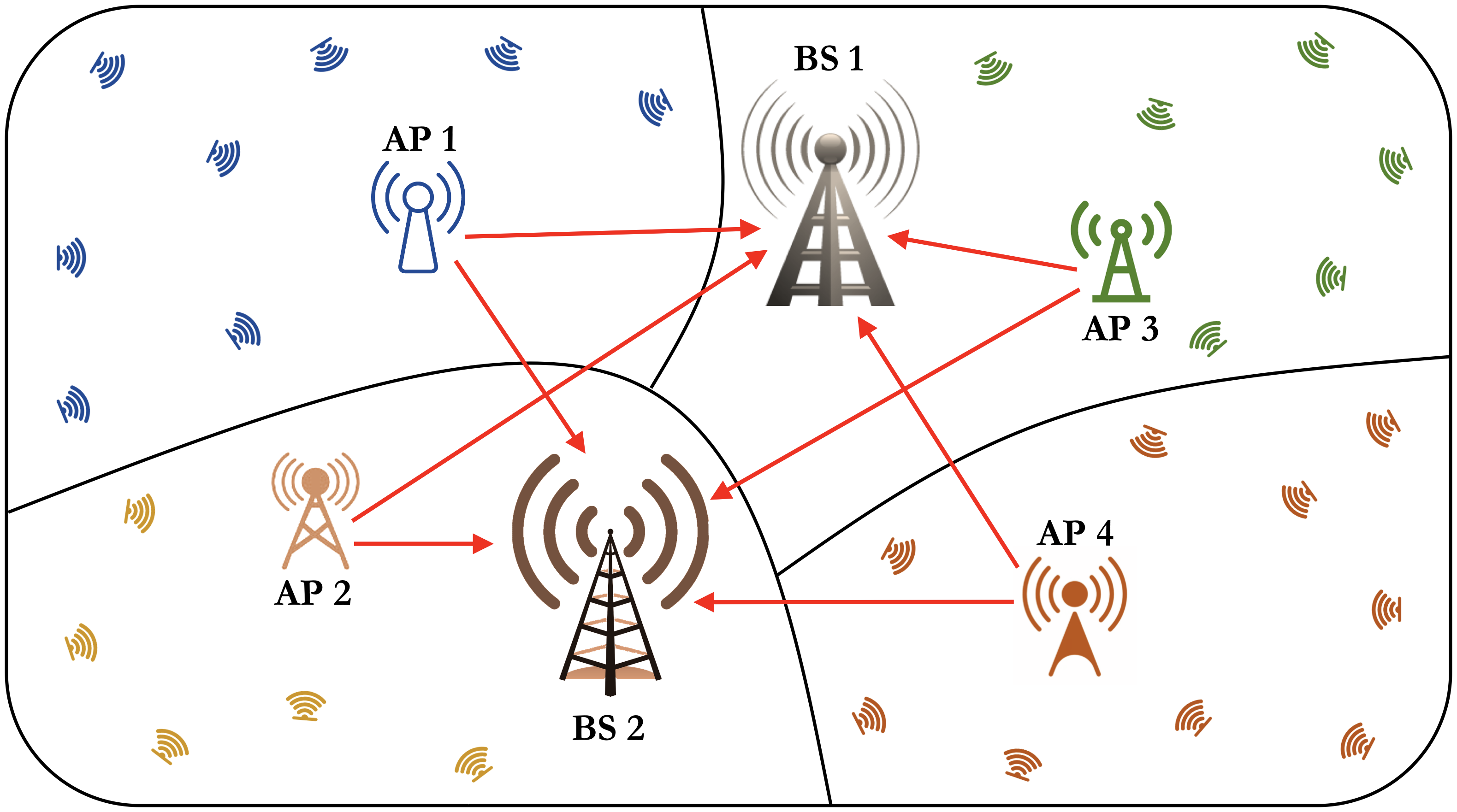}
\captionsetup{justification=justified}
\vspace{3mm}
\caption{{\small The system model and network architecture.
}}
\label{system_model_figure}
\end{figure}

We denote the collective deployment of APs and BSs by $\mathbf{P} = \left(p_1,\cdots,p_N \right)\in \mathbb{R}^{N\times 2}$ and $\mathbf{Q} = \left(q_1,\cdots,q_M \right)\in \mathbb{R}^{M\times 2}$ in which $p_n$ and $q_m$ denote the location of AP $n$ and BS $m$, respectively. In this paper, we assume that each sensor transmits its data to just one AP. As depicted in Fig. \ref{system_model_figure}, the RoI $\Omega$ is partitioned into $N$ disjoint regions $\mathbf{W} = (W_1,\cdots,W_N)\subseteq \Omega^N$ such that for each $n\in\{1,\cdots,N\}$, sensors within $W_n\subseteq \Omega$ transmit their data to AP $n$.

Our network can be regarded as a directed bipartite graph with $N+M$ vertices representing $N$ access points and $M$ base stations. Each edge from AP $n$ to BS $m$ is associated with the bit-rate $F_{n,m} \geq 0$ (bits/s) denoting the flow of data from AP $n$ to BS $m$. Thus, the routing protocol by which data is transmitted from sensors to base stations can be characterized by the flow matrix $\mathbf{F} = \big [F_{n,m}\big]_{N\times M}$ representing the flow of data between each pair of AP and BS nodes. Since the in-flow to each AP, say $n$, is equal to its out-flow, it directly follows that $R_b \int_{W_n}f(\omega)d\omega = \sum_{m=1}^{M}F_{n,m}$. This indicates that for a given cell partitioning $\mathbf{W}$, the flow matrix $\mathbf{F}$ can be uniquely determined by the normalized flow matrix $\mathbf{R} = \big [r_{n,m}\big ]_{N\times M}$ where $r_{n,m} = \frac{F_{n,m}}{\sum_{m=1}^{M}F_{n,m}}$, i.e., $r_{n,m}$ is the ratio of out-flow from AP $n$ that is transmitted to BS $m$. In particular, we have $0\leq r_{n,m} \leq 1$ and $\sum_{m=1}^M r_{n,m} = 1$ for each $n\in \{1,\cdots,N\}$.

Throughout this paper, we consider a Rayleigh flat-fading channel in which the bandwidth of the signal is small compared to the coherence bandwidth of the channel. We also assume that the receiver can track the fading process, i.e., coherent reception and the transmitter has no knowledge of the channel realization except for its statistical properties.  Our goal is to find the optimal transmission power for sensors and APs such that the allocated transmission power to each wireless link can, on average, allow the flow of data in that link to pass through\footnote{We only consider the network's communication power consumption as our objective function since the energy dissipation for transceiver electronics is negligible compared to the communication energy \cite{razzaque2014energy}.}. For this purpose, we marginalize the stochasticity of the channel due to the fading process and consider the ergodic capacity for all wireless links.

The ergodic capacity for a Rayleigh flat-fading channel between AP $n$ and BS $m$ with an average received SNR $\gamma_{n,m}$ is given by the following closed-form formula \cite{shin2003closed}:
\begin{equation}\label{ergodic_capacity_definition}
    C_{\textrm{erg}} = B \log_2(e)\times e^{\frac{1}{\gamma_{n,m}}} \times E_1\Big(\frac{1}{\gamma_{n,m}}\Big),
\end{equation}
where $E_1(z) = \int_1^\infty \frac{e^{-zx}}{x} dx$, for $\textrm{Re}\{z\} > 0$, is the exponential integral of order $1$. We consider the free-space path loss in which 
\begin{align}\label{received_SNR_gamma_ij}
    \gamma_{n,m} = \frac{P^{(n,m)}_{\small\textrm{receive}}}{\sigma B} = P^{(n,m)}_{\small\textrm{transmit}} \times \frac{G_{t_n} G_{r_m} \lambda_c^2}{\sigma B \left(4\pi\right)^2 \|p_n - q_m \|^2 },
\end{align}
where $\sigma$ and $B$ are spectral width and channel noise density, respectively, $P^{(n,m)}_{\small\textrm{transmit}}$ is AP $n$'s transmission power, $P^{(n,m)}_{\small\textrm{receive}}$ is BS $m$'s receive power, $G_{t_n}$ is the transmitter antenna gain of AP $n$, $G_{r_m}$ is the receiver antenna gain of BS $m$, and $\lambda_c$ is the wavelength of the carrier signal. By setting $C_{\textrm{erg}} = F_{n,m}$, it readily follows from Eqs. (\ref{ergodic_capacity_definition}) and (\ref{received_SNR_gamma_ij}) that the required transmission power at AP $n$ that, on average, allows the signal to be broadcasted to BS $m$ with bit-rate $F_{n,m}$ bits/s is
\begin{equation}\label{ergodic_power_AP}
    P^{(n,m)}_{\small\textrm{transmit}} = \frac{\sigma B \left(4\pi\right)^2 \|p_n - q_m\|^2 }{G_{t_n} G_{r_m} \lambda_c^2 \times U^{-1}\Big(\frac{F_{n,m}}{B\log_2(e)}\Big)},
\end{equation}
where $U(x) = e^x \times E_1(x)$. Similarly, for a sensor positioned at $\omega$, the required transmission power that, on average, allows transmission of signal to AP $n$ with bit-rate $R_b$ is given by
\begin{equation}\label{ergodic_power_sensor}
    P^{(\omega, n)}_{\small\textrm{transmit}} = \frac{\sigma B \left(4\pi\right)^2 \|p_n - \omega\|^2 }{G_{t_\textrm{sensor}} G_{r_n} \lambda_c^2 \times U^{-1}\Big(\frac{R_b}{B\log_2(e)}\Big)},
\end{equation}
where $G_{t_\textrm{sensor}}$ is the homogeneous sensors' common transmitter antenna gain. Thus, the total wireless power consumption under the ergodic capacity assumption is 
\begin{equation}\label{secondary_objective_function}
    \mathcal{P}\!\left(\mathbf{P}, \mathbf{Q}, \mathbf{W}, \mathbf{R}\right) \!=\! \sum_{n=1}^{N}\! \int_{W_n}\!\!\!\! P^{(\omega, n)}_{\small\textrm{transmit}} f(\omega)d\omega + \lambda \! \sum_{n=1}^{N}\!\sum_{m=1}^{M}\! P^{(n,m)}_{\small\textrm{transmit}},
\end{equation}
where $\lambda  \geq 0$ is the Lagrangian multiplier that provides a trade-off between sensors' and APs' transmission power consumption. Our goal in this paper is to minimize the objective function in Eq. (\ref{secondary_objective_function}) over node deployments $\mathbf{P}$, $\mathbf{Q}$, cell partitioning $\mathbf{W}$, and normalized flow matrix $\mathbf{R}$.

\section{Optimal Deployment in Rayleigh Fading WSNs}\label{Optimal_Deployment_under_Ergodic_Capacity_Assumption}

In this section, we aim at minimizing the wireless transmission power consumption $\mathcal{P}$ in Eq. (\ref{secondary_objective_function}) over node deployment $\mathbf{P}$ and $\mathbf{Q}$, cell partitioning $\mathbf{W}$, and normalized flow matrix $\mathbf{R}$. To accomplish this goal, first, we derive the necessary conditions for an optimal deployment, cell partitioning, and data routing. Then, these necessary conditions are embedded into an iterative algorithm to yield such an optimal deployment. This approach is carried out in the following three steps.

\noindent{\bf Step 1 [optimizing $\mathbf{P}$ and $\mathbf{Q}$ while $\mathbf{W}$ and $\mathbf{R}$ are fixed]: }
Using Eqs. (\ref{ergodic_power_AP}) and (\ref{ergodic_power_sensor}), we can rewrite the wireless communication power consumption $\mathcal{P}$ in Eq. (\ref{secondary_objective_function}) as 
\begin{align}\label{rewrie_second_objective_function}
    \mathcal{P}\left(\mathbf{P},\mathbf{Q},\mathbf{W},\mathbf{R} \right) = \sum_{n=1}^{N}\int_{W_n} \frac{a_n \|p_n - \omega \|^2}{U^{-1}\Big(\frac{R_b}{B\log_2(e)} \Big)}  f(\omega)d\omega + \lambda  \sum_{n=1}^{N}\sum_{m=1}^{M}\frac{b_{n,m} \|p_n - q_m \|^2}{U^{-1}\Big(\frac{F_{n,m}}{B\log_2(e)}  \Big)},
\end{align}
where $a_n = \frac{\sigma  B \times \left(4\pi\right)^2 }{G_{t_\textrm{sensor}} \times G_{r_n} \times  \lambda_c^2}$ and $b_{n,m} = \frac{\sigma  B \times \left(4\pi\right)^2}{G_{t_n} \times G_{r_m} \times  \lambda_c^2}$. For a given cell partitioning and data routing, the necessary conditions for an optimal deployment are provided next.
\begin{prop}\label{necessary_condition_deployment_ergodic_capacity}
For a fixed cell partitioning $\mathbf{W}$ and normalized flow matrix $\mathbf{R}$, the necessary conditions for the optimal AP and BS deployment $\mathbf{P}^*$ and $\mathbf{Q}^*$ in a heterogeneous Rayleigh fading WSN with wireless transmission power consumption defined in Eq. (\ref{secondary_objective_function}) are given by:
\begin{align}\label{necessary_condition_ergodic_AP}
    p_n^* &= \frac{\frac{a_n v_n }{U^{-1}\big(\frac{R_b}{B\log_2(e)} \big)}\!\times\! c_n + \lambda  \sum\limits_{m=1}^{M}\frac{b_{n,m}}{U^{-1}\big(\frac{F_{n,m}}{B\log_2(e)}  \big)}\!\times\! q_m^* }{\frac{a_n v_n}{U^{-1}\big(\frac{R_b}{B\log_2(e)} \big)} + \lambda \sum\limits_{m=1}^{M}\frac{b_{n,m}}{U^{-1}\big(\frac{F_{n,m}}{B\log_2(e)}  \big)}   },  \\
    q^*_m &=\frac{\sum\limits_{n=1}^{N}\frac{b_{n,m}}{U^{-1}\big(\frac{F_{n,m}}{B\log_2(e)}  \big)}\times p^*_n }{\sum\limits_{n=1}^{N}\frac{b_{n,m}}{U^{-1}\big(\frac{F_{n,m}}{B\log_2(e)}  \big)}}, \label{necessary_condition_ergodic_BS}
\end{align}
for all $n\in \mathcal{I}_{AP} = \{1,\cdots,N\}$ and $m\in \mathcal{I}_{BS} = \{1,\cdots,M\}$, where $v_n$ and $c_n$ are the volume and centroid of $W_n$, respectively. 

The proof of Proposition \ref{necessary_condition_deployment_ergodic_capacity} is provided in Appendix \ref{proof_of_ergodic_node_necessary_condition}.
\end{prop}

\noindent{\bf Step 2 [optimizing $\mathbf{W}$ while $\mathbf{P}$, $\mathbf{Q}$, and $\mathbf{R}$ are fixed]: }
The flow matrix $\mathbf{F}$'s dependence on cell partitioning is only through  individual cells' volumes. This indicates that the AP transmission power consumption in Eq. (\ref{rewrie_second_objective_function}) only depends on the volume of regions $W_1,\cdots,W_N$ and not their shape. This is in contrast to sensors' power consumption in Eq. (\ref{rewrie_second_objective_function}) which depends on both volume and shape of each cell. This observation implies that by holding the volumes constant, we can adjust the region boundaries to reduce the sensors' power consumption while the APs' power consumption is kept fixed. Using this intuition, we obtain the following property of the optimal region boundaries.
\begin{lemma}\label{ergodic_property_of_region_boundaries}
Let $\mathbf{W}^* = \left(W^*_1, \cdots, W^*_N\right)$ be an optimal cell partitioning that minimizes the network's transmission power consumption $\mathcal{P}$ in Eq. (\ref{rewrie_second_objective_function}) for a given node deployment $\mathbf{P}$, $\mathbf{Q}$, and normalized flow matrix $\mathbf{R}$. Let $\delta^*_{i,j} = W^*_i \cap W^*_j$ be the optimal boundary between neighboring regions $W^*_i$ and $W^*_j$. If $a_i = a_j$, then $\delta^*_{i,j}$ is a segment; otherwise, $\delta^*_{i,j}$ is an arc with its center placed at $c=\frac{a_i p_i - a_j p_j}{a_i - a_j}$.
\end{lemma}

The proof of Lemma \ref{ergodic_property_of_region_boundaries} is provided in Appendix \ref{proof_of_ergodic_boundaries}.

Assume that the optimal boundary $\delta^*_{i,j}$ in Lemma \ref{ergodic_property_of_region_boundaries} intersects the line $\overline{p_i p_j}$ at the point $h^*_{i,j}$. The following proposition provides a necessary condition on $h^*_{i,j}$'s position.

\begin{prop}\label{ergodic_necessary_optimal_cell_partitioning}
For a given node deployment $\mathbf{P}$, $\mathbf{Q}$, and data routing $\mathbf{R}$, let $\mathbf{W}^*=\left(W_1^*,\cdots,W_N^* \right)$ be an optimal cell partitioning that minimizes the objective function $\mathcal{P}$ in Eq. (\ref{rewrie_second_objective_function}). Let $\delta^*_{i,j} = W^*_i \cap W^*_j$ be the boundary between neighboring cells $W^*_i$ and $W^*_j$ that intersects the line $\overline{p_i p_j}$ at $h^*_{i,j}$. We have
\begin{align}\label{ergodic_optimal_cells_eq}
    &\textrm{ }\textrm{ }\textrm{ }\textrm{ } \frac{a_i}{U^{-1}\Big(\frac{R_b}{B\log_2(e)} \Big)} \|p_i - h^*_{i,j}\|^2 + \lambda \sum_{t=1}^{M}\frac{b_{i,t} \|p_i-q_t\|^2 \times \frac{r_{i,t}R_b}{B\log_2(e)}}{U^{-1}\Big(\frac{r_{i,t}R_bv^*_i}{B\log_2(e)}\Big)\Big[1 - \frac{r_{i,t}R_bv^*_i}{B\log_2(e)} \times U^{-1}\Big(\frac{r_{i,t}R_bv^*_i}{B\log_2(e)}\Big) \Big]} \nonumber\\
   &= \frac{a_j}{U^{-1}\Big(\frac{R_b}{B\log_2(e)} \Big)} \|p_j - h^*_{i,j}\|^2 + \lambda \sum_{t=1}^{M}\frac{b_{j,t} \|p_j-q_t\|^2\times\frac{r_{j,t}R_b}{B\log_2(e)}}{U^{-1}\Big(\frac{r_{j,t}R_bv^*_j}{B\log_2(e)}\Big)\Big[1 - \frac{r_{j,t}R_bv^*_j}{B\log_2(e)} \times U^{-1}\Big(\frac{r_{j,t}R_bv^*_j}{B\log_2(e)}\Big) \Big]}.    
\end{align}

The proof of Proposition \ref{ergodic_necessary_optimal_cell_partitioning} is 
provided in Appendix \ref{proof_of_ergodic_cell_partitioning}.
\end{prop}

\noindent{\bf Step 3 [optimizing $\mathbf{R}$ while $\mathbf{P}$, $\mathbf{Q}$, and $\mathbf{W}$ are fixed]: }
The choice of normalized flow matrix $\mathbf{R}$ does not affect the sensors' power consumption and only changes the APs' power consumption in Eq. (\ref{rewrie_second_objective_function}). Since the cell partitioning, and thus  each individual cell's volume is fixed, the total amount of data per time unit that each AP transmits is fixed. Therefore, to optimize $\mathbf{R}$, each AP can work independent of other AP nodes to adjust its data transmission scheme. For each AP, say $n$, we have the following objective function:
\begin{align}\label{N_objective_functions_ergodic}
    &\argmin_{\substack{F_{n,1} \cdots  F_{n,M}}} \sum_{m=1}^{M} \frac{b_{n,m}}{U^{-1}\Big(\frac{F_{n,m}}{B\log_2(e)}  \Big)} \|p_n - q_m \|^2, \\    &\textrm{s.t. } \sum_{m=1}^{M} F_{n,m} = \int_{W_n} R_b f(\omega)d\omega = R_b v_n, \label{N_objective_functions_ergodic_eq1} \\
    &\qquad  F_{n,m} \geq 0 \textrm{ for all } m\in \mathcal{I}_{BS}. \label{N_objective_functions_ergodic_eq2}
\end{align}
To make the above constrained optimization problem tractable, we make use of a common optimization technique that aims to minimize an upper bound of the objective function instead of explicitly minimizing the objective function itself. Thus, our first goal is to present an upper bound on AP $n$'s transmission power consumption in Eq. (\ref{N_objective_functions_ergodic}). \begin{lemma}\label{upper_bound_on_U}
Let $U(x)=e^x \times E_1(x)$ where $E_1(x)$ is the exponential integral of order $1$. Then, we have
\begin{equation}\label{upper_bound_on_U_Eq}
 e^x - 1 <  \frac{1}{U^{-1}(x)} < \frac{e^{2x}-1}{2}.
\end{equation}

The proof of Lemma \ref{upper_bound_on_U} is provided in Appendix \ref{proof_of_upper_bound_on_U}.
\end{lemma}

Lemma \ref{upper_bound_on_U} yields the following upper bound on AP $n$'s power consumption in Eq. (\ref{N_objective_functions_ergodic}):
\begin{align}\label{upper_bound_N_objective_functions_ergodic}
    &\argmin_{\substack{F_{n,1} \cdots  F_{n,M}}} \sum_{m=1}^{M} b_{n,m} \|p_n - q_m \|^2\times \frac{e^{\frac{2F_{n,m}}{B\log_2(e)}} - 1}{2}, \\
    &\textrm{s.t. } \sum_{m=1}^{M} F_{n,m} = \int_{W_n} R_b f(\omega)d\omega = R_b v_n, \label{upper_bound_N_objective_functions_ergodic_eq1} \\
    &\qquad  F_{n,m} \geq 0 \textrm{ for all } m\in \mathcal{I}_{BS}. \label{upper_bound_N_objective_functions_ergodic_eq2}
\end{align}

A systematic approach is presented in Algorithm \ref{ergodic_capacity_optimize_Q_while_P_and_W_are_fixed_algorithm} to provide the optimal solution to the constrained optimization problem in Eqs. (\ref{upper_bound_N_objective_functions_ergodic})-(\ref{upper_bound_N_objective_functions_ergodic_eq2}).

\begin{algorithm}[ht!]
\SetAlgoLined
\SetKwRepeat{Do}{do}{while}
\centering
\includegraphics[width=130mm]{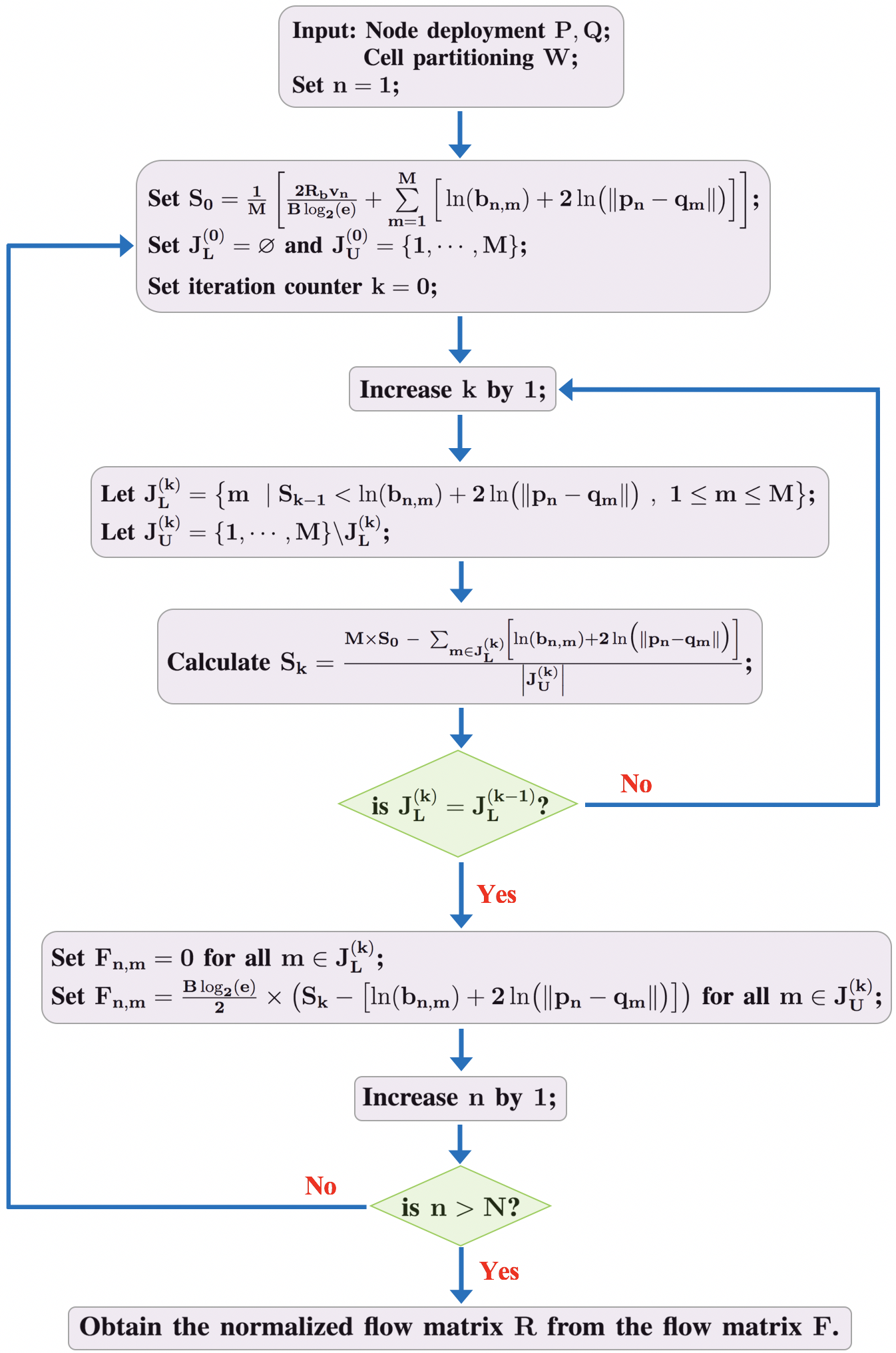}
\caption{Optimal data routing in two-tier WSNs under ergodic capacity assumption}
\label{ergodic_capacity_optimize_Q_while_P_and_W_are_fixed_algorithm}
\end{algorithm}

\begin{prop}\label{optimal_routing_for_ergodic_capacity}
For a given node deployment $\mathbf{P}$, $\mathbf{Q}$, and cell partitioning $\mathbf{W}$, Algorithm \ref{ergodic_capacity_optimize_Q_while_P_and_W_are_fixed_algorithm} provides the optimal solution to the constrained optimization problem in Eqs. (\ref{upper_bound_N_objective_functions_ergodic})-(\ref{upper_bound_N_objective_functions_ergodic_eq2}) and returns the optimal normalized flow matrix $\mathbf{R}$.

The proof of Proposition \ref{optimal_routing_for_ergodic_capacity} is provided in Appendix \ref{proof_of_ergodic_optimal_routing}.
\end{prop}

Next, we present Algorithm \ref{PEEL_Algorithm}, named  Power-Efficient Ergodic-based Lloyd (PEEL) Algorithm, that makes use of the properties we explored in this section to minimize the wireless transmission power consumption $\mathcal{P}$ in Eq. (\ref{rewrie_second_objective_function}) over node deployment, cell partitioning, and data routing.

\begin{algorithm}[ht!]
\SetAlgoLined
\SetKwRepeat{Do}{do}{while}
\centering
\includegraphics[width=102mm]{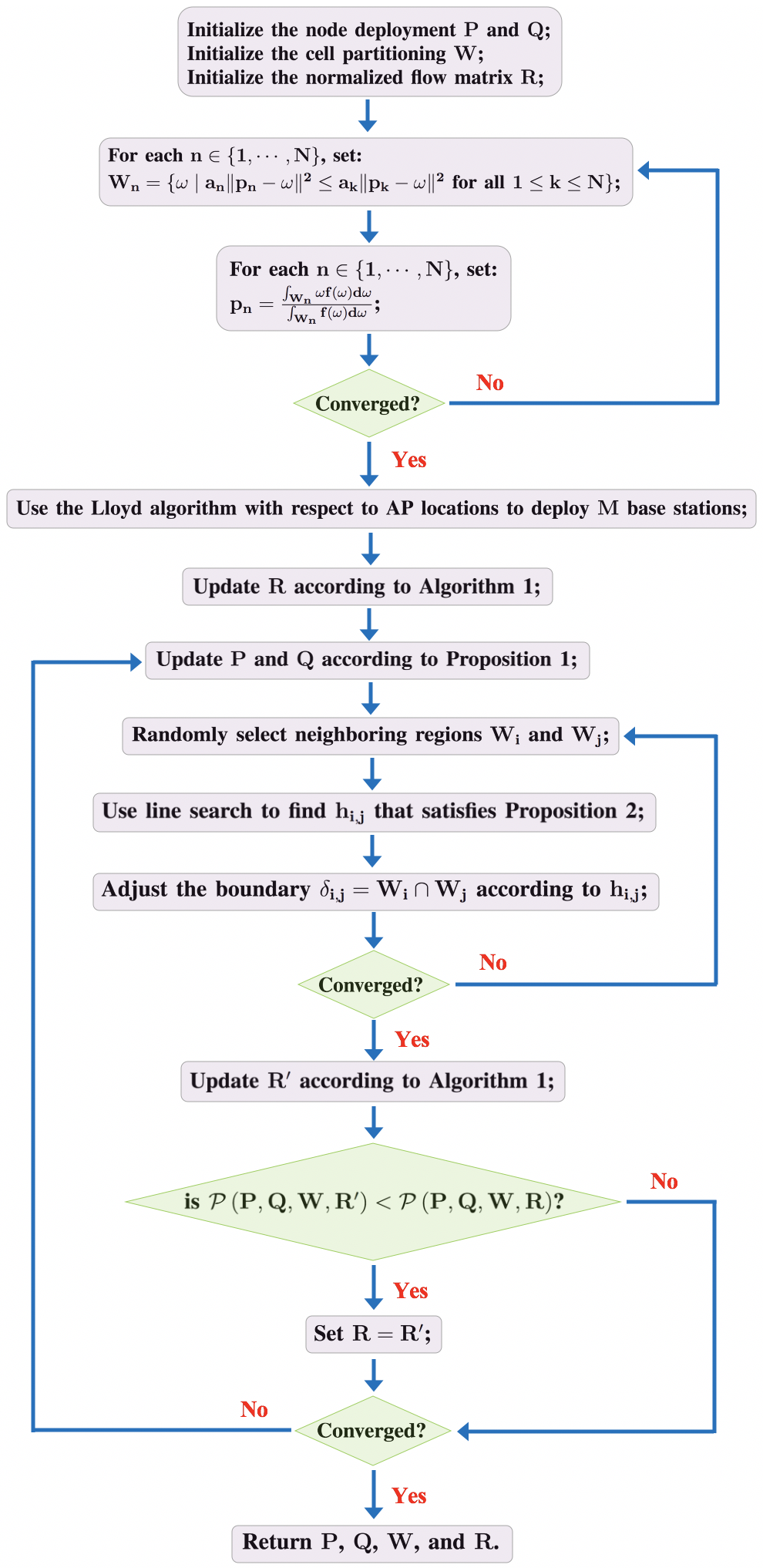}
\caption{Power-Efficient Ergodic-based Lloyd (PEEL) Algorithm}
\label{PEEL_Algorithm}
\end{algorithm}

\begin{prop}\label{convergence_PEEL_algorithm}
The PEEL algorithm is an iterative improvement algorithm and converges.
\end{prop}

The proof of Proposition \ref{convergence_PEEL_algorithm} is provided in Appendix \ref{proof_of_PEEL_convergence}.

\section{Experiments}\label{experiments}

We carry out simulations for a heterogeneous Rayleigh fading WSN consisting of $3$ BSs, $15$ APs, and $1000$ sensors. 
The region of interest $\Omega$ is a square field of size $10\textrm{km}\times 10\textrm{km}$. For all homogeneous sensors,  the bit-rate and transmitter antenna gain are $R_b = 30$Kbps and $G_{t_\textrm{sensor}} = 1$, respectively, and the carrier wavelength is $\lambda_c = 3$m. We denote BS $m$'s receiver antenna gain by $G^{\textrm{(BS)}}_{r_m}$ and AP $n$'s transmitter and receiver antenna gains by $G^{\textrm{(AP)}}_{t_n}$ and $G^{\textrm{(AP)}}_{r_n}$, respectively. These values are set as follows:

\begin{align}\label{antenna_gains}
G^{\textrm{(AP)}}_{t_n}&=
    \begin{cases}
      2 & \text{if}\ n\in \{1,2,3,4,8,9,10 \} \\
      4 &  \textrm{otherwise,}
    \end{cases} \nonumber\\  
G^{\textrm{(AP)}}_{r_n}&=
    \begin{cases}
      2 & \text{if}\ n\in \{1,2,5,6,8,9,12,13 \} \\
      4 & \textrm{otherwise,}
    \end{cases} \nonumber \\
G^{\textrm{(BS)}}_{r_m}&=
    \begin{cases}
      2 & \text{if}\ m\in \{1,2\} \\
      4 & \textrm{otherwise.}
    \end{cases}   
\end{align}

We consider a spectral width and noise density of $B=500$KHz and $\sigma=2\times 10^{-17}$ Watts/Hz for all communication channels, respectively. Using the aforementioned experimental setup, we can calculate the coefficients $a_n$ and $b_{n,m}$ in Eq. (\ref{rewrie_second_objective_function}) accordingly. For example, we have $b_{8,1} = \frac{\sigma  B \times \left(4\pi\right)^2}{G^{\textrm{(AP)}}_{t_8} \times G^{\textrm{(BS)}}_{r_1} \times  \lambda_c^2} \simeq 4.39 \times 10^{-11}$ Watts/$\textrm{m}^2$.  The Lagrangian multiplier in the objective function is set to $\lambda  = 0.25$. Finally, we carry on the simulations for two different sensor density functions: (i) Uniform PDF; and (ii) Mixture of Gaussian PDF defined as:
\begin{align*}
    f(\omega) &= \frac{1}{2}  \times  \mathcal{N} \left( \begin{bmatrix}
    3\times 10^3 \\
    3\times 10^3 
\end{bmatrix} , \begin{bmatrix}
    1.5\times 10^6 & 0 \\
    0 & 1.5\times 10^6
\end{bmatrix}  \right) \\& +  \frac{1}{4}  \times  \mathcal{N} \left( \begin{bmatrix}
    6\times 10^3 \\
    7\times 10^3 
\end{bmatrix} , \begin{bmatrix}
    2\times 10^6 & 0 \\
    0 & 2\times 10^6 
\end{bmatrix} \right) \\& +  \frac{1}{4}  \times  \mathcal{N} \left( \begin{bmatrix}
    7.5\times 10^3 \\
    2.5\times 10^3 
\end{bmatrix} , \begin{bmatrix}
    10^6 & 0 \\
    0 & 10^6
\end{bmatrix}  \right).
\end{align*}

We compare our proposed PEEL Algorithm with cluster formation (CF) Algorithm \cite{chatterjee2015multiple}, heterogeneous two-tier Lloyd (HTTL) Algorithm \cite{karimi2020energy}, particle swarm optimization (PSO) Algorithm \cite{dandekar2013energy}, and virtual force (VFA) Algorithm \cite{zou2004sensor}. The primary incentive for choosing these algorithms is that they represent state-of-the-art methods in different categories used by researchers for deploying ndodes in WSNs. The CF algorithm falls within the category of methods that take a graph-theoretic approach for load balancing and energy efficiency. The HTTL algorithm belongs to the family of geometric-based methods in which the target region is partitioned into several regions, one for each network node, based on a predefined measure of closeness. The PSO algorithm represents the class of meta-heuristic node deployment techniques in which optimization tools are used to find optimal node positions. Finally, the VFA algorithm is a prominent example of force-based techniques and has inspired numerous methods that  apply virtual forces to relocate nodes.

The weighted transmission power consumption of the heterogeneous WSN outlined above is summarized in Table \ref{weighted_power_ergodic_capacity} for the CF, HTTL, PEEL, PSO, and VFA algorithms. Since our objective function considers the ergodic capacity for all communication channels, these results can  be interpreted as the amount of power that can, on average, allow the flow of data in each link to pass through. The PEEL algorithm 
outperforms all other algorithms for both uniform and Gaussian mixture sensor density functions. In particular, the PEEL algorithm reduces the power consumption of the second best algorithm by more than a factor of $2$. This in turn prolongs the network lifetime and leads to a more sustainable network architecture.

\begin{table}[!bth]
\centering
\caption{Weighted power consumption comparison $(\mathcal{P})$}
\begin{tabular}{|cc||ccccccccc|}
 \toprule
               {\bf Method}                                    &\!\!\!\!\!&    CF    &\!\!\!\!\!&   HTTL   &\!\!\!\!\!&     {\bf PEEL}   &\!\!\!\!\!&   PSO   &\!\!\!\!\!&   VFA    \\\hline \hline
{\bf {\small Weighted Power (mW) for the uniform PDF}}         &\!\!\!\!\!& $177.56$ &\!\!\!\!\!& $17.94$  &\!\!\!\!\!& $\mathbf{8.61}$  &\!\!\!\!\!& $96.37$ &\!\!\!\!\!& $22.62$  \\ 
{\bf {\small Weighted Power (mW) for the Gaussian mixture PDF}}&\!\!\!\!\!& $41.65$  &\!\!\!\!\!& $14.60$  &\!\!\!\!\!& $\mathbf{6.76}$  &\!\!\!\!\!& $49.41$ &\!\!\!\!\!& $52.19$  \\ 
 \bottomrule 
\end{tabular}
  \label{weighted_power_ergodic_capacity}
\end{table}

\begin{figure}
\centering
\includegraphics[width=80mm]{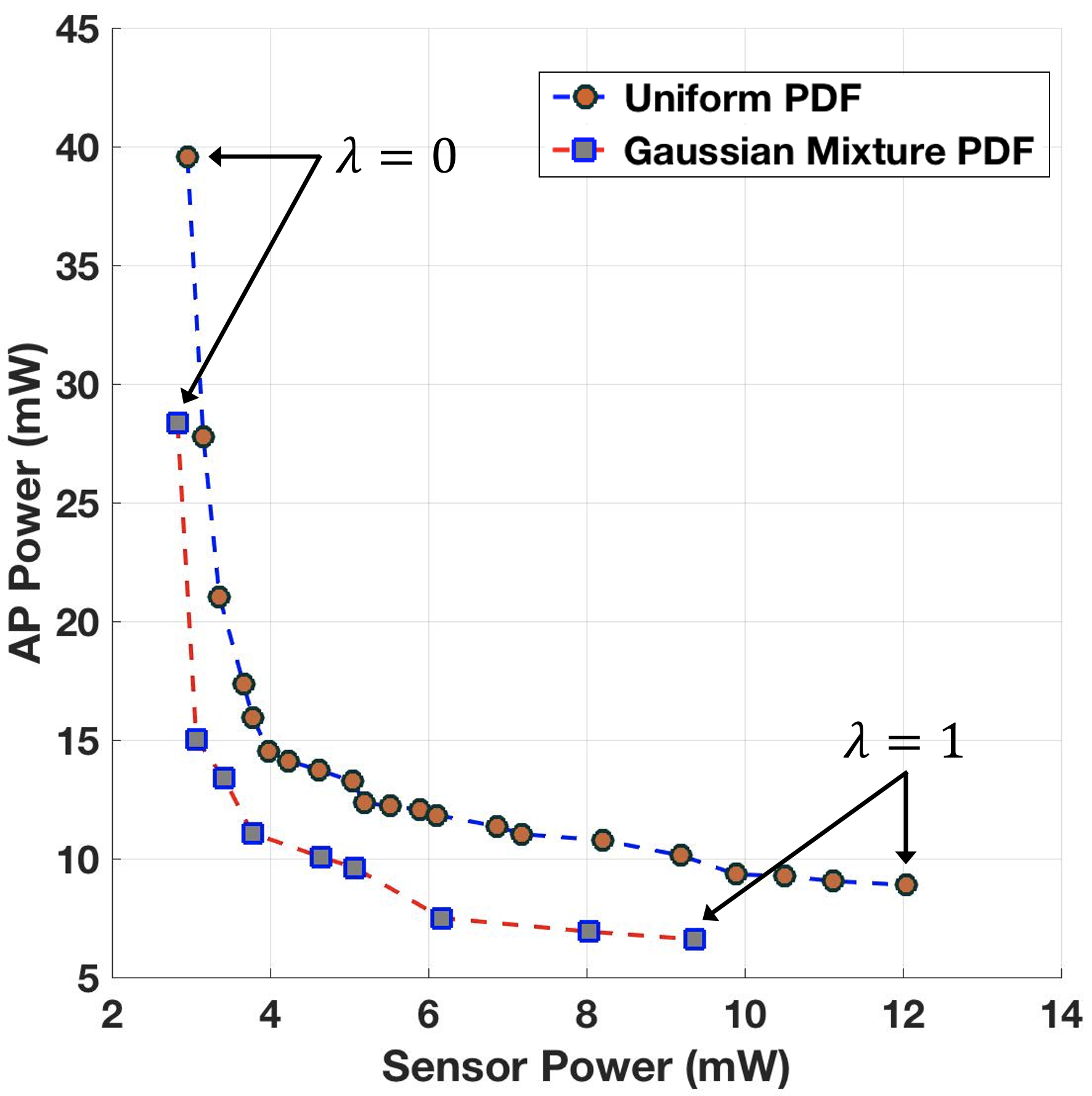}
\captionsetup{justification=justified}
\caption{\small{AP-Sensor power trade-off for the PEEL algorithm for the uniform and Gaussian mixture sensor density functions.}}
\label{effect_of_beta}
\end{figure}

Many factors contribute to the superior performance of the PEEL algorithm. According to the Shannon's capacity formula, the required SNR at the receiver is an exponential function of the transmission bit-rate for an error-free information retrieval; however, the majority of methods in the literature consider a linear approximation to this exponential relationship. This in turn leads to an underestimation of the actual power consumption of the network. On the contrary, 
we exploited the exponential relationship between SNR and bit-rate and designed our node deployment algorithm accordingly. Another major contributing factor is that this exponential relationship between the required SNR and data flow is exploited in Algorithm \ref{ergodic_capacity_optimize_Q_while_P_and_W_are_fixed_algorithm} to find the optimal routing for data transfer from sensors to base stations.

Finally, we study the trade-off between sensors' and APs' power consumption in Eq. (\ref{secondary_objective_function}) that is parameterized by the Lagrangian multiplier $\lambda$. For small values of $\lambda $, the objective function $\mathcal{P}$ is dominated by the sensor power consumption; hence, minimizing the sensors' power consumption is more important than APs' power consumption. 
However, the importance of minimizing APs' power consumption increases as $\lambda$ increases. This behavior is illustrated in Fig. \ref{effect_of_beta} for the uniform and Gaussian mixture sensor density functions as $\lambda$ is increased from $0$ to $1$ for the same initial node deployment. In general, a Gaussian Mixture distribution requires less power since the sensors are concentrated around three locations and can be covered with smaller average distances. Fig. \ref{effect_of_beta} shows that increasing $\lambda$ decreases the APs' power consumption but increases sensors' power consumption. This is in line with Eq. (\ref{necessary_condition_ergodic_AP}) since access points tend to become closer to base stations and farther away from centroids and sensors as $\lambda$ increases.

\section{Conclusion}\label{conclusion}

We studied a heterogeneous Rayleigh fading sensor network comprised of sensors that monitor the environment, access points that act as relays, and base stations where all sensory data are transmitted to by the means of wireless communication. Our goal is to minimize the network's wireless transmission power consumption that incorporates both large-scale and small-scale propagation effects. By considering the ergodic capacity for all communication channels, we derive the theoretical necessary conditions for the optimal deployment, cell partitioning, and data routing. Not only is the network's power consumption minimized, but also the allocated transmission power to each channel can, on average, allow the flow of data in that channel to pass through. An iterative algorithm is then proposed accordingly to deploy nodes. Simulation results demonstrated the superiority of our proposed algorithm over similar methods in the literature and established its efficacy in reducing the communication power consumption of such networks.

\appendices

\section{Proof of Proposition \ref{necessary_condition_deployment_ergodic_capacity}}\label{proof_of_ergodic_node_necessary_condition}

Using the parallel axis theorem, we can rewrite the objective function $\mathcal{P}$ in Eq. (\ref{rewrie_second_objective_function}) as
\begin{align}\label{Appendix_F_Eq1}
    \mathcal{P}\left(\mathbf{P},\mathbf{Q},\mathbf{W},\mathbf{R} \right) = &\sum_{n=1}^{N}\int_{W_n} \frac{a_n}{U^{-1}\Big(\frac{R_b}{B\log_2(e)} \Big)} \|c_n - \omega \|^2 f(\omega)d\omega \nonumber\\&+
    \sum_{n=1}^{N} \frac{a_n}{U^{-1}\Big(\frac{R_b}{B\log_2(e)} \Big)} \|p_n - c_n \|^2 v_n \nonumber\\&+
    \lambda  \sum_{i=1}^{N}\sum_{j=1}^{M}\frac{b_{i,j}}{U^{-1}\Big(\frac{F_{i,j}}{B\log_2(e)}  \Big)} \|p_i - q_j \|^2.
\end{align}
Since the optimal deployment satisfies the zero-gradient equation, for each $i\in \mathcal{I}_{AP}$, we have
\begin{align}\label{Appendix_F_Eq2}
    \frac{\partial}{\partial p^*_i}\mathcal{P} = \frac{2 a_i}{U^{-1}\Big(\frac{R_b}{B\log_2(e)} \Big)} \big(p^*_i - c_i\big) v_i + \lambda  \sum_{j=1}^{M}\frac{2 b_{i,j}}{U^{-1}\Big(\frac{F_{i,j}}{B\log_2(e)}  \Big)} \big(p^*_i - q^*_j\big) = 0.
\end{align}
Similarly, by taking the derivative with respect to the location of BS $i\in \mathcal{I}_{BS}$, we have
\begin{align}\label{Appendix_F_Eq3}
    \frac{\partial}{\partial q^*_i}\mathcal{P} = \lambda  \sum_{j=1}^{N}\frac{2 b_{j,i}}{U^{-1}\Big(\frac{F_{j,i}}{B\log_2(e)}  \Big)} \big(q^*_i - p^*_j \big) = 0.
\end{align}
By solving these two equations, we obtain Eqs. (\ref{necessary_condition_ergodic_AP}) and (\ref{necessary_condition_ergodic_BS}) and the proof is complete. $\hfill\blacksquare$

\section{Proof of Lemma \ref{ergodic_property_of_region_boundaries}}\label{proof_of_ergodic_boundaries}

First, we prove the following lemma.

\begin{lemma}\label{difference_in_weighted_square_distances}
For a constant $d\in \mathbb{R}$, the geometric locus of points $\omega\in \mathbb{R}^2$ that satisfy the equation
\begin{equation}\label{proof_lemma1_Eq1}
    a_i \|p_i - \omega\|^2 - a_j\|p_j - \omega\|^2= d,
\end{equation}
is a line perpendicular to $\overline{p_ip_j}$ in the case of $a_i=a_j$, and either  a circle centered at $c=\frac{a_ip_i - a_jp_j}{a_i - a_j}$ or an empty set in the case of $a_i\neq a_j$.
\end{lemma}

\textit{Proof:} First, we consider the case where $a_i = a_j = a$. Let $h$ be the projection of the point $\omega$ on the line $\overline{p_ip_j}$. Using Pythagoras' theorem, we can rewrite Eq. (\ref{proof_lemma1_Eq1}) as follows:
\begin{equation}\label{proof_lemma1_Eq0.5}
    \left(\|p_i-h\|^2 + \|h-\omega\|^2 \right) - \left(\|p_j-h\|^2 + \|h-\omega\|^2 \right) = \left(\|p_i-h\|^2 - \|p_j-h\|^2 \right) = \frac{d}{a},
\end{equation}
thus, any point $\omega$ whose projection on the line $\overline{p_ip_j}$ is $h$ satisfies Eq. (\ref{proof_lemma1_Eq1}). Therefore, the geometric locus of the point $\omega$ is a line perpendicular to the line $\overline{p_ip_j}$. Now, we consider the case where $a_i \neq a_j$. Let $p = (p_x, p_y)$ and $\omega = (\omega_x, \omega_y)$. We can rewrite Eq. (\ref{proof_lemma1_Eq1}) as:
\begin{align}\label{proof_lemma1_Eq2}
    (a_i \!-\! a_j) \!\left(\omega_x^2 + \omega_y^2 \right) - 2\left(a_i p_{ix} \!-\! a_j p_{jx} \right) \omega_x - 2\left(a_i p_{iy} \!-\! a_j p_{jy} \right) \omega_y = d \!-\! \left(a_i \|p_i\|^2 - a_j\|p_j\|^2 \right),  
\end{align}
or equivalently:
\begin{align}\label{proof_lemma1_Eq3}
    \left[\omega_x - \frac{a_ip_{ix} - a_jp_{jx}}{a_i - a_j}  \right]^2 + \left[\omega_y - \frac{a_ip_{iy} - a_jp_{jy}}{a_i - a_j}  \right]^2 = d',
\end{align}
where $d' = \frac{d - \left(a_i \|p_i\|^2 - a_j\|p_j\|^2\right)}{a_i - a_j} + \frac{\left(a_ip_{ix} - a_jp_{jx}\right)^2 + \left(a_ip_{iy} - a_jp_{jy} \right)^2}{(a_i - a_j)^2}$. Hence, the geometric locus of the point $\omega$ is either an empty set if $d' < 0$ or a circle centered at $c=\frac{a_ip_i - a_jp_j}{a_i - a_j}$ with radius $\kappa=\sqrt{d'}$ and Lemma \ref{difference_in_weighted_square_distances} is proved. $\hfill\blacksquare$

Now, we use proof by contradiction to establish Lemma \ref{ergodic_property_of_region_boundaries}. Let $v^*_i$ and $v^*_j$ be the volume of the neighboring regions $W^*_i$ and $W^*_j$, respectively, and assume that the optimal boundary $\delta^*_{i,j}$ is neither a segment if $a_i=a_j$, nor an arc when $a_i \neq a_j$. Let $m_{i,j}(\alpha) = \alpha p_i + (1-\alpha)p_j$ and let $l_{i,j}(\alpha)$ be either a line perpendicular to $\overline{p_ip_j}$ at $m_{i,j}(\alpha)$ in case $a_i=a_j$, or a circle centered at $c_{i,j}=\frac{a_ip_i - a_jp_j}{a_i-a_j}$ and radius $\kappa_{i,j}(\alpha) = \|c_{i,j} - m_{i,j}(\alpha)\|$ in case $a_i \neq a_j$. Now, we define:
\begin{align}\label{proof_lemma1_Eq4}
    W'_i &= \left\{\omega \mid \omega\in \Omega^*_{i,j}, a_i \|p_i \!-\! \omega\|^2 - a_j\|p_j\!-\!\omega\|^2 \leq a_i \|p_i \!-\! m_{i,j}(\alpha)\|^2 - a_j\|p_j\!-\!m_{i,j}(\alpha)\|^2 \right\}, \\
    W'_j &= \left\{\omega \mid \omega\in \Omega^*_{i,j}, a_i \|p_i \!-\! \omega\|^2 - a_j\|p_j\!-\!\omega\|^2 \geq a_i \|p_i \!-\! m_{i,j}(\alpha)\|^2 - a_j\|p_j\!-\!m_{i,j}(\alpha)\|^2 \right\},    \label{proof_lemma1_Eq5}
\end{align}
where $\Omega^*_{i,j} = W^*_i \cup W^*_j$, and let $v'_i(\alpha)$ and $v'_j(\alpha)$ be the volume of regions $W'_i$ and $W'_j$, respectively. Note that since the sensor density function $f(\omega)$ is a continuous and differentiable function, both $v'_i(\alpha)$ and $v'_j(\alpha)$ are continuous functions of $\alpha$. As an intuition, note that the geometric locus of points $\omega\in \mathbb{R}^2$, such that  $a_i \|p_i \!-\! \omega\|^2 - a_j\|p_j\!-\!\omega\|^2 = a_i \|p_i \!-\! m_{i,j}(\alpha)\|^2 - a_j\|p_j\!-\!m_{i,j}(\alpha)\|^2$ holds, is $l_{i,j}(\alpha)$ according to Lemma \ref{difference_in_weighted_square_distances}. It readily follows from simple geometric reasoning that for $a_i \leq a_j$, we have $v'_i = 0$ for large enough values of $\alpha$, and $v'_j = 0$ for $\alpha = \frac{a_i}{a_i - a_j}$ that leads to $\kappa_{i,j}(\alpha)=0$. Similarly, for $a_i > a_j$, we have $v'_i = 0$ for $\alpha = \frac{a_i}{a_i - a_j}$ that leads to $\kappa_{i,j}(\alpha)=0$, and $v'_j=0$ for large enough values of $\alpha$.

Using the above argument and the fact that $v'_i(\alpha) + v'_j(\alpha) = v^*_i + v^*_j$, it readily follows that there exists an $\alpha^*$ for which we have  $v'_i(\alpha^*) = v^*_i$ and $v'_j(\alpha^*) = v^*_j$. Now, we define a new cell partitioning $\mathbf{W}'' = \left(W''_1, \cdots, W''_N \right)$ where $W''_t = W^*_t$ for $t \notin \{i,j\}$, $W''_i = W'_i(\alpha^*)$, and $W''_j = W'_j(\alpha^*)$. Then, substituting $\mathbf{W}^*$ with $\mathbf{W}''$ will increase the objective function by:
\begin{align}\label{Appendix_G_Eq1}
    \Delta &= \Bigg[\sum_{n=1}^{N}\int_{W''_n} \frac{a_n}{U^{-1}\Big(\frac{R_b}{B\log_2(e)} \Big)} \|p_n - \omega \|^2 f(\omega)d\omega + \lambda  \sum_{i=1}^{N}\sum_{j=1}^{M}\frac{b_{i,j}}{U^{-1}\Big(\frac{r_{i,j}R_b v''_i}{B\log_2(e)}  \Big)} \|p_i - q_j \|^2 \Bigg] \nonumber\\&-
    \Bigg[\sum_{n=1}^{N}\int_{W^*_n} \frac{a_n}{U^{-1}\Big(\frac{R_b}{B\log_2(e)} \Big)} \|p_n - \omega \|^2 f(\omega)d\omega + \lambda  \sum_{i=1}^{N}\sum_{j=1}^{M}\frac{b_{i,j}}{U^{-1}\Big(\frac{r_{i,j}R_b v^*_i}{B\log_2(e)}  \Big)} \|p_i - q_j \|^2 \Bigg].
\end{align}
Since $W''_t = W^*_t$ for $t\notin\{i,j\}$ and $v''_t=v^*_t$ for all $t\in \{1,\cdots,N\}$, we have:
\begin{align}\label{Appendix_G_Eq2}
    U^{-1}\bigg(\frac{R_b}{B\log_2(e)} \bigg) \times \Delta &= \Bigg[\int_{W''_i} a_i \|p_i - \omega \|^2 f(\omega)d\omega + \int_{W''_j} a_j \|p_j - \omega \|^2 f(\omega)d\omega  \Bigg] \nonumber\\&-
    \Bigg[\int_{W^*_i} a_i \|p_i - \omega \|^2 f(\omega)d\omega + \int_{W^*_j} a_j \|p_j - \omega \|^2 f(\omega)d\omega  \Bigg].    
\end{align}
Let $\mathcal{V}_1 = W''_i \cap W^*_j$ and $\mathcal{V}_2 = W''_j \cap W^*_i$. Note that both $\mathcal{V}_1$ and $\mathcal{V}_2$ are non-empty; otherwise, we have $W''_i = W^*_i$ and $W''_j = W^*_j$ which contradicts the assumption that the optimal boundary $\delta^*_{i,j}$ is not a segment or an arc. Now, we can rewrite Eq. (\ref{Appendix_G_Eq2}) as follows:
\begin{align}
    U^{-1}\bigg(\frac{R_b}{B\log_2(e)} \bigg) \times \Delta &= \bigg [ \int_{\mathcal{V}_1} a_{i} \|p_i - \omega\|^2  f(\omega) d\omega + \int_{\mathcal{V}_2} a_{j} \|p_j - \omega\|^2  f(\omega) d\omega  \bigg] \nonumber\\&-
    \bigg [ \int_{\mathcal{V}_2} a_{i} \|p_i - \omega\|^2  f(\omega) d\omega + \int_{\mathcal{V}_1} a_{j} \|p_j - \omega\|^2  f(\omega) d\omega  \bigg] \label{proof_lemma1_Eq9}\\ 
    &=  \int_{\mathcal{V}_1} \left(a_{i} \|p_i - \omega\|^2 - a_{j} \|p_j - \omega\|^2 \right)  f(\omega) d\omega \nonumber\\& + \int_{\mathcal{V}_2} \left(a_{j} \|p_j - \omega\|^2 - a_{i} \|p_i - \omega\|^2 \right)  f(\omega) d\omega  \label{proof_lemma1_Eq10}\\
    &< \int_{\mathcal{V}_1} \left(a_{i} \big|\! \big|p_i - m_{i,j}(\alpha^*)\big |\! \big|^2 - a_{j} \big|\! \big|p_j - m_{i,j}(\alpha^*) \big|\! \big|^2 \right)  f(\omega) d\omega \nonumber \\&+ \int_{\mathcal{V}_2} \left(a_{j} \big |\! \big|p_j - m_{i,j}(\alpha^*)\big |\! \big|^2 - a_{i} \big |\! \big|p_i - m_{i,j}(\alpha^*)\big |\! \big|^2 \right)  f(\omega) d\omega \label{proof_lemma1_Eq11}\\&=
    \left(a_{i} \big|\! \big|p_i \!-\! m_{i,j}(\alpha^*)\big |\! \big|^2 - a_{j} \big|\! \big|p_j \!-\! m_{i,j}(\alpha^*) \big|\! \big|^2 \right)\!\!\times\!\! \left(\int_{\mathcal{V}_1}\!\!f(\omega)d\omega - \int_{\mathcal{V}_2}\!\!f(\omega)d\omega \!\right) \label{proof_lemma1_Eq12} \\&=0, \label{proof_lemma1_Eq13}
\end{align}
where the inequality in (\ref{proof_lemma1_Eq11}) follows from Lemma \ref{difference_in_weighted_square_distances} and the fact that both $\mathcal{V}_1$ and $\mathcal{V}_2$ are non-empty. Also, Eq. (\ref{proof_lemma1_Eq13}) follows from the fact that $\mathcal{V}_1$ and $\mathcal{V}_2$ have the same volume because $v''_i= v^*_i$ and $v''_j = v^*_j$. Thus, since $U^{-1}(x) > 0$ for all $x > 0$, we have
\begin{align}\label{Appendix_G_Eq3}
    U^{-1}\bigg(\frac{R_b}{B\log_2(e)} \bigg) \times \Delta < 0  \qquad \Longrightarrow \qquad \Delta < 0,
\end{align}
which contradicts the optimality of $\mathbf{W}^*$ and the proof is complete. $\hfill\blacksquare$

\section{Proof of Proposition \ref{ergodic_necessary_optimal_cell_partitioning}}\label{proof_of_ergodic_cell_partitioning}

As shown in Lemma \ref{ergodic_property_of_region_boundaries}, the optimal boundary $\delta^*_{i,j}$ is a segment if $a_i = a_j$, or an arc with its center located at $c = \frac{a_i p_i - a_j p_j}{a_i - a_j}$ if $a_i \neq a_j$. Let $h^*_{i,j}$ be the intersection point of $\delta^*_{i,j}$ and the line $\overline{p_i p_j}$ which corresponds to the scalar $\alpha^*$ that satisfies the equation $\alpha^* p_i + (1 - \alpha^*)p_j = h^*_{i,j}$. For a small and positive $\gamma > 0$, let $\alpha' = \alpha^* - \gamma$ and define the new cell partitioning $\mathbf{W}' = \big(W'_1, \cdots, W'_N \big)$ as $W'_t = W^*_t$ for $t\notin \{i,j\}$ and
\begin{align}\label{appendix_H_eq1}
    W'_i &= \big\{\omega \mid \omega \in \Omega^*_{i,j}, a_i \|p_i - \omega \|^2 - a_j \|p_j - \omega \|^2 \leq a_i \|p_i - h'_{i,j} \|^2 - a_j \|p_j - h'_{i,j} \|^2 \big\}, \\ 
    W'_j &= \big\{\omega \mid \omega \in \Omega^*_{i,j}, a_i \|p_i - \omega \|^2 - a_j \|p_j - \omega \|^2 \geq a_i \|p_i - h'_{i,j} \|^2 - a_j \|p_j - h'_{i,j} \|^2 \big\},   \label{appendix_H_eq2}  
\end{align}
where $\Omega^*_{i,j} = W^*_i \cup W^*_j$ and $h'_{i,j} = \alpha' p_i + (1 - \alpha') p_j$. The infinitesimal difference between $\alpha^*$ and $\alpha'$ causes infinitesimal difference between volumes of regions $W'_i$ and $W'_j$, i.e., $v'_i$ and $v'_j$, and volumes of regions $W^*_i$ and $W^*_j$. In other words, if $dv$ is the volume of the region $dW = W'_i - W^*_i = W^*_j - W'_j$, we have $v'_i = v^*_i + dv$ and $v'_j = v^*_j - dv$. The increase in the sensor power consumption due to replacing $\mathbf{W}^*$ by $\mathbf{W}'$ is then given by
\begin{align}
 \Delta_1 \!&=\! \int_{W'_i}\frac{a_i}{U^{-1}\Big(\frac{R_b}{B\log_2(e)} \Big)} \|p_i - \omega \|^2 f(\omega)d\omega +  \int_{W'_j}\frac{a_j}{U^{-1}\Big(\frac{R_b}{B\log_2(e)} \Big)} \|p_j - \omega \|^2 f(\omega)d\omega \nonumber\\&-
 \int_{W^*_i}\frac{a_i}{U^{-1}\Big(\frac{R_b}{B\log_2(e)} \Big)} \|p_i - \omega \|^2 f(\omega)d\omega -  \int_{W^*_j}\frac{a_j}{U^{-1}\Big(\frac{R_b}{B\log_2(e)} \Big)} \|p_j - \omega \|^2 f(\omega)d\omega, \label{appendix_H_eq3}
\end{align}
which can be simplified to
\begin{align}\label{appendix_H_eq4}
    \Delta_1  = \int_{dW}\frac{1}{U^{-1}\Big(\frac{R_b}{B\log_2(e)} \Big)} \Big[a_i \|p_i - \omega \|^2 - a_j \|p_j - \omega \|^2 \Big] f(\omega)d\omega.
\end{align}
Using Lemma \ref{difference_in_weighted_square_distances} and the definition of $W'_i$ and $W'_j$ in Eqs. (\ref{appendix_H_eq1}) and (\ref{appendix_H_eq2}), it readily follows that
\begin{align}\label{appendix_H_eq5}
    \Delta_1 = \frac{1}{U^{-1}\Big(\frac{R_b}{B\log_2(e)} \Big)} \Big[a_i \|p_i - h^*_{i,j} \|^2 - a_j \|p_j - h^*_{i,j} \|^2 \Big] dv + \mathcal{O}(dv^2),
\end{align}
where $\mathcal{O}(dv^2)$ includes terms of second and higher order. The increase in AP power consumption due to substituting $\mathbf{W}^*$ with $\mathbf{W}'$ can be written as
\begin{align}\label{appendix_H_eq6}
    \Delta_2 &= \sum_{t=1}^{M}\frac{b_{i,t}}{U^{-1}\Big(\frac{r_{i,t}R_bv'_i}{B\log_2(e)}  \Big)} \|p_i - q_t \|^2 + \sum_{t=1}^{M}\frac{b_{j,t}}{U^{-1}\Big(\frac{r_{j,t}R_bv'_j}{B\log_2(e)}  \Big)} \|p_j - q_t \|^2 \nonumber \\&-
    \sum_{t=1}^{M}\frac{b_{i,t}}{U^{-1}\Big(\frac{r_{i,t}R_bv^*_i}{B\log_2(e)}  \Big)} \|p_i - q_t \|^2 - \sum_{t=1}^{M}\frac{b_{j,t}}{U^{-1}\Big(\frac{r_{j,t}R_bv^*_j}{B\log_2(e)}  \Big)} \|p_j - q_t \|^2.
\end{align}
For the function $U(x) = e^x E_1(x)$, we have $\frac{d}{dx}U(x) = e^xE_1(x)+e^x\times\frac{-e^{-x}}{x} = U(x) - \frac{1}{x}$ and
\begin{align}\label{appendix_H_eq7}
    \frac{d}{dx}\bigg(\frac{1}{U^{-1}(x)}\bigg) = \frac{-\big(U^{-1}\big)'(x)}{\big[U^{-1}(x)\big]^2} = \frac{-1}{\big[U^{-1}(x)\big]^2 \times U'\big(U^{-1}(x)\big)} = \frac{1}{U^{-1}(x)\big[1 - x U^{-1}(x)\big]}.
\end{align}
Now, using Taylor series expansion, we can rewrite $\Delta_2$ in Eq. (\ref{appendix_H_eq6}) as follows:
\begin{align}
    \Delta_2 &= \sum_{t=1}^{M}\Bigg[\frac{b_{i,t}\|p_i - q_t\|^2 \times \frac{r_{i,t}R_b}{B\log_2(e)}}{U^{-1}\Big(\frac{r_{i,t}R_bv^*_i}{B\log_2(e)}\Big)\times \Big[1 - \frac{r_{i,t}R_bv^*_i}{B\log_2(e)} \times U^{-1}\Big(\frac{r_{i,t}R_bv^*_i}{B\log_2(e)}\Big)\Big]}\Bigg] dv \nonumber \\ &-
    \sum_{t=1}^{M}\Bigg[\frac{b_{j,t}\|p_j - q_t\|^2 \times \frac{r_{j,t}R_b}{B\log_2(e)}}{U^{-1}\Big(\frac{r_{j,t}R_bv^*_j}{B\log_2(e)}\Big)\times \Big[1 - \frac{r_{j,t}R_bv^*_j}{B\log_2(e)} \times U^{-1}\Big(\frac{r_{j,t}R_bv^*_j}{B\log_2(e)}\Big)\Big]}\Bigg] dv + \mathcal{O}(dv^2), \label{appendix_H_eq8}
\end{align}
where $\mathcal{O}(dv^2)$ contains terms of second and higher order in Taylor series expansion. Thus, the total increase in the objective function is given by $\Delta = \Delta_1 + \lambda  \Delta_2$. Since the cell partitioning $\mathbf{W}^*$ is optimal, the increase in the objective function due to replacing $\mathbf{W}^*$ by $\mathbf{W}'$ cannot be negative; thus, we have $\Delta \geq 0$. Therefore, by dividing $\Delta$ by $dv > 0$ and taking the limit $dv \longrightarrow 0$, the term $\mathcal{O}(dv^2)$ vanishes and we have
\begin{align}
    \frac{a_i}{U^{-1}\Big(\frac{R_b}{B\log_2(e)} \Big)} \|p_i - h^*_{i,j}\|^2 + \lambda \sum_{t=1}^{M}\frac{b_{i,t} \|p_i-q_t\|^2\times\frac{r_{i,t}R_b}{B\log_2(e)}}{U^{-1}\Big(\frac{r_{i,t}R_bv^*_i}{B\log_2(e)}\Big)\Big[1 - \frac{r_{i,t}R_bv^*_i}{B\log_2(e)} \times U^{-1}\Big(\frac{r_{i,t}R_bv^*_i}{B\log_2(e)}\Big) \Big]} \nonumber\\
   \geq \frac{a_j}{U^{-1}\Big(\frac{R_b}{B\log_2(e)} \Big)} \|p_j - h^*_{i,j}\|^2 + \lambda \sum_{t=1}^{M}\frac{b_{j,t} \|p_j-q_t\|^2\times\frac{r_{j,t}R_b}{B\log_2(e)}}{U^{-1}\Big(\frac{r_{j,t}R_bv^*_j}{B\log_2(e)}\Big)\Big[1 - \frac{r_{j,t}R_bv^*_j}{B\log_2(e)} \times U^{-1}\Big(\frac{r_{j,t}R_bv^*_j}{B\log_2(e)}\Big) \Big]}.   \label{appendix_H_eq9} 
\end{align}
Now, by defining $\alpha'' = \alpha^* + \gamma$ for an infinitesimal $\gamma > 0$ and repeating the same argument, we obtain $\Delta \leq 0$ and the inequality sign in Eq. (\ref{appendix_H_eq9}) will be reversed, i.e., we have
\begin{align}
    \frac{a_i}{U^{-1}\Big(\frac{R_b}{B\log_2(e)} \Big)} \|p_i - h^*_{i,j}\|^2 + \lambda \sum_{t=1}^{M}\frac{b_{i,t} \|p_i-q_t\|^2\times\frac{r_{i,t}R_b}{B\log_2(e)}}{U^{-1}\Big(\frac{r_{i,t}R_bv^*_i}{B\log_2(e)}\Big)\Big[1 - \frac{r_{i,t}R_bv^*_i}{B\log_2(e)} \times U^{-1}\Big(\frac{r_{i,t}R_bv^*_i}{B\log_2(e)}\Big) \Big]} \nonumber\\
   \leq \frac{a_j}{U^{-1}\Big(\frac{R_b}{B\log_2(e)} \Big)} \|p_j - h^*_{i,j}\|^2 + \lambda \sum_{t=1}^{M}\frac{b_{j,t} \|p_j-q_t\|^2\times\frac{r_{j,t}R_b}{B\log_2(e)}}{U^{-1}\Big(\frac{r_{j,t}R_bv^*_j}{B\log_2(e)}\Big)\Big[1 - \frac{r_{j,t}R_bv^*_j}{B\log_2(e)} \times U^{-1}\Big(\frac{r_{j,t}R_bv^*_j}{B\log_2(e)}\Big) \Big]}.   \label{appendix_H_eq10} 
\end{align}
The two inequalities in Eqs. (\ref{appendix_H_eq9}) and (\ref{appendix_H_eq10}) yield Eq. (\ref{ergodic_optimal_cells_eq}) and the proof is complete. $\hfill\blacksquare$

\section{Proof of Lemma \ref{upper_bound_on_U}}\label{proof_of_upper_bound_on_U}

First, we show that the function $U(x)$ is strictly decreasing. For this purpose, we have:
\begin{align}\label{lemma3_proof_eq1}
    \frac{d}{dx}U(x) = e^x E_1(x) + e^x\times\Big(-\frac{e^{-x}}{x}\Big) = e^x E_1(x)- \frac{1}{x} < 0,
\end{align}
where the last inequality follows from the inequality $xe^xE_1(x)<1$ in \cite{nantomah2021some}. Note that for $y > 0$, the function $U(y)=e^yE_1(y)$ satisfies the following inequalities \cite{nantomah2021some, gautschi1959some}:
\begin{align}\label{lemma3_proof_eq2}
    \frac{1}{2}\ln \Big(1 + \frac{2}{y}\Big) < U(y) < \ln \Big(1 + \frac{1}{y}\Big).
\end{align}
Eq. (\ref{lemma3_proof_eq2}) shows that both domain and range of the function $U(y)$ is $(0,\infty)$. Since $U(y)$ is strictly decreasing, it is invertible and we define $y = U^{-1}(x)$. 
Substituting $y = U^{-1}(x)$ and $x=U(y)$ in Eq. (\ref{lemma3_proof_eq2}), we obtain:
\begin{align}\label{lemma3_proof_eq3}
    \frac{1}{2}\ln \Big(1 + \frac{2}{U^{-1}(x)}\Big) < x < \ln \Big(1 + \frac{1}{U^{-1}(x)}\Big).
\end{align}
From Eq. (\ref{lemma3_proof_eq3}), we have
\begin{align}\label{lemma3_proof_eq4}
     \frac{1}{2}\ln \Big(1 + \frac{2}{U^{-1}(x)}\Big) < x \qquad &\Longrightarrow \qquad \frac{1}{U^{-1}(x)} < \frac{e^{2x}-1}{2}, \\
     x < \ln \Big(1 + \frac{1}{U^{-1}(x)}\Big) \qquad &\Longrightarrow \qquad e^x - 1 < \frac{1}{U^{-1}(x)},
\end{align}
which concludes the proof. $\hfill\blacksquare$

\section{Proof of Proposition \ref{optimal_routing_for_ergodic_capacity}}\label{proof_of_ergodic_optimal_routing}

First, we prove the following lemma.
\begin{lemma}\label{Appendix_D_Eq1}
Let $g(x) = a^x + a^{C-x}$ where $x\in [0, C]$ for $a,C\in \mathbb{R}^+$ and $a>1$. Then, $g(.)$ is symmetric around the point $x=\frac{C}{2}$ and  strictly decreasing in the interval $\big[0, \frac{C}{2} \big)$.
\end{lemma}

\textit{Proof: } The function $g(.)$ is symmetric because $g(x) = g(C-x)$. Now, by taking the derivative w.r.t. $x$, we have $\frac{d}{dx}g(x) = \ln(a)\times \big(a^x - a^{C-x} \big)$. Since $a > 1$, we have $\frac{d}{dx}g(x) < 0$ for $x\in \big[0, \frac{C}{2} \big)$ and the proof is complete. $\hfill\blacksquare$

Lemma \ref{Appendix_D_Eq1} leads to the following conclusion.

\begin{corollary}\label{Appendix_D_Eq2}
Let $x_1$ and $x_2$ be two non-negative real numbers such that $x_1 + x_2 = C$ is a constant. Then, for $a>1$, decreasing $|x_1 - x_2|$ results in smaller $a^{x_1} + a^{x_2}$ values. 
\end{corollary}
We can rewrite the constrained objective function formulation in Eqs. (\ref{upper_bound_N_objective_functions_ergodic})-(\ref{upper_bound_N_objective_functions_ergodic_eq2}) as
\begin{align}\label{Appendix_J_Eq1}
    &\argmin_{F_{n,1},\cdots,F_{n,M}} \sum_{i=1}^{M} e^{\Big[\frac{2F_{n,i}}{B\log_2(e)} + \ln \big(b_{n,i} \|p_n - q_i \|^2\big) \Big]}, \\
    \textrm{s.t. } \sum_{i=1}^{M} F_{n,i} = \int_{W_n} &R_b f(\omega)d\omega = R_b v_n, \qquad \textrm{and} \qquad F_{n,i} \geq 0 \textrm{ for all } i\in \mathcal{I}_{BS}. \label{Appendix_J_Eq2} 
\end{align}
The above constrained objective function formulation can be rewritten as
\begin{align}\label{Appendix_J_Eq3}
    &\argmin_{x_{n,1},\cdots,x_{n,M}} \sum_{i=1}^{M} e^{x_{n,i}} \\
    &\textrm{s.t. } \sum_{i=1}^{M} x_{n,i} = \frac{2R_b v_n}{B\log_2(e)} + \sum_{i=1}^{M}\ln \big(b_{n,i} \|p_n - q_i \|^2\big) = S, \label{Appendix_J_Eq4}\\
    &\qquad x_{n,i} \geq \ln \big(b_{n,i} \|p_n - q_i \|^2\big) \qquad \textrm{for all } i\in \{1,\cdots,M\},    \label{Appendix_J_Eq5}
\end{align}
where $x_{n,i} = \frac{2F_{n,i}}{B\log_2(e)} + \ln \big(b_{n,i} \|p_n - q_i \|^2\big)$. According to Corollary \ref{Appendix_D_Eq2}, the objective function in Eq. (\ref{Appendix_J_Eq3}) can be decreased by reducing $| x_{n,i} - x_{n,j} |$ while keeping their summation fixed. Hence, the minimum occurs when we have $x_{n,1} = \cdots = x_{n,M} = \frac{S}{M}$. However, it is crucial to make sure that the constraints in Eqs. (\ref{Appendix_J_Eq4}) and (\ref{Appendix_J_Eq5}) are not contradicted. We can always improve the objective function in Eq. (\ref{Appendix_J_Eq3}) and achieve a lower value by decreasing the distance between any pair of $x_{n,i}$ and $x_{n,j}$ while keeping their summation constant as long as the constraints in Eq. (\ref{Appendix_J_Eq5}) are not contradicted. This observation results in the following corollary.
\begin{corollary}\label{Appendix_J_Eq6}
Let $X^*_n = (x^*_{n,1}, \cdots, x^*_{n,M})$ be the optimal solution to the constrained objective function in Eqs. (\ref{Appendix_J_Eq3})$-$(\ref{Appendix_J_Eq5}). Then, there exist unique sets $J^*_L$ and $J^*_U$ such that
\begin{align}\label{Appendix_J_Eq7}
    x^*_{n,i} = x^*_{n,j} = \overline{x}^* \textrm{ } \textrm{ for } \textrm{ } \forall i,j \in J^*_U, \quad\textrm{ and }\quad
    x^*_{n,i} = \ln\big( b_{n,i}\| p_n - q_i  \|^2 \big) \textrm{ } \textrm{ for } \textrm{ } \forall i \in J^*_L,
\end{align}
and $x^*_{n,i} > \overline{x}^*$ for all $i \in J^*_L$.
\end{corollary}
To see why the last property holds, first, let us assume that we have $x^*_{n,j} = \ln\big( b_{n,j}\| p_n - q_j  \|^2 \big)$ for all $j\in J^*_U$. Since $v_n > 0$, it follows that
\begin{equation}\label{Appendix_D_Eq11}
    \sum_{t=1}^{M}x^*_{n,t} = \sum_{t\in J^*_U}x^*_{n,t} + \sum_{t\in J^*_L}x^*_{n,t} = \sum_{t=1}^{M} \ln\big( b_{n,t}\| p_n - q_t  \|^2 \big) < S,
\end{equation}
which is in contradiction with Eq. (\ref{Appendix_J_Eq4}). Hence, there exists an index $j'\in J^*_U$ for which $x^*_{n,j'} > \ln\big( b_{n,j'}\| p_n - q_{j'}  \|^2 \big)$. Now, assume that there exists an index $i$ such that $x^*_{n,i} = \ln\big( b_{n,i}\| p_n - q_i  \|^2 \big) < \overline{x}^* = x^*_{n,j'}$. Then, according to Corollary \ref{Appendix_D_Eq2}, we can achieve a lower objective function by replacing $x^*_{n,i}$ and $x^*_{n,j'}$ with $x^*_{n,i} + \eta$ and $x^*_{n,j'} - \eta$ for any $0 < \eta < x^*_{n,j'} - \ln\big( b_{n,j'}\| p_n - q_{j'}  \|^2 \big)$, which contradicts the optimality of $X^*$. Thus, we have $x^*_{n,i} > \overline{x}^*$ for all $i \in J^*_L$. 

Corollary \ref{Appendix_J_Eq6} indicates that in an optimal solution, all $x^*_{n,t}$ values should be equal to some value $\overline{x}^*$ except for those that cannot get close enough to $\overline{x}^*$ without contradicting Eq. (\ref{Appendix_J_Eq5}). Hence, the optimal solution can be found using a water filling algorithm as follows. By initializing $J_L$ to an empty set and starting from the case in which all $x_{n,t}$ values are equal to the mean value $\overline{x} = \frac{S}{M}$, we can identify those indices such as $i\in J$ for which $x_{n,i} < \ln\big( b_{n,i}\| p_n - q_i  \|^2 \big)$. Thus, $J$ provides the first series of indices for which the value of $x_{n,i}$ cannot be reduced enough to the mean value $\overline{x}$ without contradicting the constraint in Eq. (\ref{Appendix_J_Eq5}). Therefore, the optimal value for each $i\in J$ is $x^*_{n,i} = \ln\big( b_{n,i}\| p_n - q_i  \|^2 \big)$ and we update the set $J_L$ by taking its union with the set $J$. Now, we can update the mean value $\overline{x}$ such that $\sum_{i\in \mathcal{I}_{BS}\backslash J_L}\overline{x} + \sum_{i\in J_L}x^*_{n,i}$ or equivalently $\big(M - |J_L|\big)\times \overline{x} + \sum_{i\in J_L} \ln\big( b_{n,i}\| p_n - q_i  \|^2 \big)$ still sums to $S$. By using the new mean value $\overline{x}$, we can determine the next series of indices that would belong to $J_L$ and the same procedure can be repeated. Note that in each iteration, the mean value $\overline{x}$ either decreases or stays the same and the set $J_L$ either increases in size or stays the same. If $J_L$ stays the same, meaning that there has been no other index that would contradict Eq. (\ref{Appendix_J_Eq5}), then we have found the optimal solution and the algorithm terminates. Since $|J_L| \leq M$, the process of $J_L$ increasing in size can continue for at most $M$ iterations and the algorithm will finally converge to the optimal value $X_n^*$ that satisfies Eq. (\ref{Appendix_J_Eq7}) in Corollary \ref{Appendix_J_Eq6}. The above procedure is summarized in Algorithm \ref{ergodic_capacity_optimize_Q_while_P_and_W_are_fixed_algorithm}. Note that the optimal values $F^*_{n,1}, \cdots, F^*_{n,M}$ in Eqs. (\ref{Appendix_J_Eq1}) and (\ref{Appendix_J_Eq2}) can then be found as $F^*_{n,i} = \frac{B\log_2(e)}{2} \times \big[x^*_{n,i} -\ln\big( b_{n,i}\| p_n - q_i  \|^2 \big)\big]$ and the proof is complete. $\hfill\blacksquare$

\section{Proof of Proposition \ref{convergence_PEEL_algorithm}}\label{proof_of_PEEL_convergence}

First, we aim to prove the convergence of the initialization step that is outlined in Algorithm \ref{PEEL_Algorithm}. Note that the generalized Voronoi diagram $\mathcal{V}=(V_1,\cdots,V_N)$, in which $V_n = \{\omega \mid a_n\|p_n-\omega \|^2 \leq a_k \|p_k - \omega\|^2 \textrm{ for all } 1 \leq k \leq N \}$,  provides the optimal cell partitioning for the following cost function:
\begin{align}\label{Appendix_E_Eq1}
    \mathcal{D}(\mathbf{P}, \mathbf{W}) = \sum_{n=1}^{N}\int_{W_n} a_n \|p_n - \omega \|^2 f(\omega)d\omega.
\end{align}
Thus, for a fixed AP deployment $\mathbf{P}$, updating $\mathbf{W}$ according to $\mathcal{V}$ in Algorithm \ref{PEEL_Algorithm} does not increase the cost function $\mathcal{D}$. Now, using the parallel axis theorem, we can rewrite Eq. (\ref{Appendix_E_Eq1}) as follows:
\begin{align}\label{Appendix_E_Eq2}
    \mathcal{D}(\mathbf{P}, \mathbf{W}) = \sum_{n=1}^{N}\int_{W_n} a_n \|p_n - c_n \|^2 f(\omega)d\omega + \sum_{n=1}^{N}\int_{W_n} a_n \|c_n - \omega \|^2 f(\omega)d\omega.
\end{align}
Hence, for a fixed cell partitioning $\mathbf{W}$, updating $\mathbf{P}$ according to the rule $p_n=c_n = \frac{\int_{W_n}\omega f(\omega)d\omega}{\int_{W_n}f(\omega)d\omega}$ does not increase the cost function $\mathcal{D}$ in Eq. (\ref{Appendix_E_Eq2}) either. Therefore, by iterating this process, a sequence of non-increasing $\mathcal{D}$ values is generated and since $\mathcal{D}\geq 0$, it will converge.

Note that base stations are initialized by applying the Lloyd algorithm to the set of AP points, which is known to converge. Finally, the normalized flow matrix $\mathbf{R}$ is updated by applying Algorithm \ref{ergodic_capacity_optimize_Q_while_P_and_W_are_fixed_algorithm} and converges as shown in Appendix \ref{proof_of_ergodic_optimal_routing}. Thus, the initialization step which is outlined in Algorithm \ref{PEEL_Algorithm} will eventually converge.

Now, to establish the convergence of the PEEL algorithm, we demonstrate that none of the three steps in the PEEL algorithm will increase the objective function $\mathcal{P}$. Note that the term in Eq. (\ref{necessary_condition_ergodic_AP}) is the solution to the zero-gradient equation; thus, for a fixed $\mathbf{W}$, $\mathbf{R}$, $\mathbf{Q}$, and $\{p_j\}_{j\neq i}$, updating $p_i$ according to Eq. (\ref{necessary_condition_ergodic_AP}) does not increase the objective function. Similarly, updating $q_i$ according to Eq. (\ref{necessary_condition_ergodic_BS}) does not increase the objective function. Therefore, updating node deployment according to the PEEL Algorithm will not increase $\mathcal{P}$. Note that cell partitioning is updated iteratively in the PEEL algorithm as follows: in each iteration, two adjacent regions are selected and their common boundary is adjusted according to the optimal necessary condition; thus, the objective function either remains the same or decreases at each iteration. In the last step, the optimal routing that minimizes the upper bound on AP power consumption is calculated. If the resulting routing leads to a decrease in the original objective function, it will be preserved; otherwise, it will be discarded. Hence, the objective function does not increase as a result of the data routing update rule. Hence, Algorithm \ref{PEEL_Algorithm} generates a non-increasing sequence of $\mathcal{P}$ values, which proves its convergence since $\mathcal{P}$ is lower-bounded $0$. $\hfill\blacksquare$

\bibliographystyle{ieeetr}
\bibliography{reference}

\end{document}